\documentclass[sigconf, authorversion]{acmart}

\AtBeginDocument{%
  \providecommand\BibTeX{{%
    \normalfont B\kern-0.5em{\scshape i\kern-0.25em b}\kern-0.8em\TeX}}}

\copyrightyear{2022} 
\acmYear{2022} 
\setcopyright{acmlicensed}\acmConference[CUI 2022]{4th Conference on Conversational User Interfaces}{July 26--28, 2022}{Glasgow, United Kingdom}
\acmBooktitle{4th Conference on Conversational User Interfaces (CUI 2022), July 26--28, 2022, Glasgow, United Kingdom}
\acmPrice{15.00}
\acmDOI{10.1145/3543829.3544528}
\acmISBN{978-1-4503-9739-1/22/07}

\acmBooktitle{} 

\usepackage{color, colortbl}
\usepackage{multirow}
\usepackage{subfig}
\usepackage{hyperref}
\usepackage[font=small,labelfont=bf]{caption}

\begin{document}

\title{Rules Of Engagement: Levelling Up To Combat Unethical CUI Design}

\author{Thomas Mildner}
\affiliation{%
  \institution{University of Bremen}
  \city{Bremen}
  \country{Germany}
}
\email{mildner@uni-bremen.de}
\orcid{0000-0002-1712-0741 }

\author{Gian-Luca Savino}
\affiliation{%
  \institution{University of St.Gallen}
  \city{St.Gallen}
  \country{Switzerland}
}
\email{gian-luca.savino@unisg.ch}
\orcid{0000-0002-1233-234X}

\author{Philip Doyle}
\affiliation{%
  \institution{University College Dublin}
  \city{Dublin}
  \country{Ireland}
}
\email{philip.doyle1@ucdconnect.ie}
\orcid{0000-0002-2686-8962}

\author{Rainer Malaka}
\affiliation{%
  \institution{University of Bremen}
  \city{Bremen}
  \country{Gremany}
}
\email{malaka@tzi.de}
\orcid{0000-0001-6463-4828}

\renewcommand{\shortauthors}{Mildner et al.}

\begin{abstract}
While a central goal of HCI has always been to create and develop interfaces that are easy to use, a deeper focus has been set more recently on designing interfaces more ethically. However, the exact meaning and measurement of ethical design has yet to be established both within the CUI community and among HCI researchers more broadly. In this provocation paper we propose a simplified methodology to assess interfaces based on five dimensions taken from prior research on so-called dark patterns. As a result, our approach offers a numeric score to its users representing the manipulative nature of evaluated interfaces. It is hoped that the approach - which draws a distinction between persuasion and manipulative design, and focuses on how the latter functions rather than how it manifests - will provide a viable way for quantifying instances of unethical interface design that will prove useful to researchers, regulators and potentially even users.
\end{abstract}

\ccsdesc[500]{Human-centered computing~User interface design}
\ccsdesc[500]{Human-centered computing~User centered design}
\ccsdesc[300]{Human-centered computing~HCI theory, concepts and models}

\keywords{ethical design, dark patterns, conversational user interfaces, evaluation methods and techniques}


\maketitle

\section{Introduction}
Although ethical interface design has been a concern in HCI for some time, researchers are only beginning to formulate ways to understand, define and measure occurrences of unethical interface design. In terms of attempts to build knowledge around this issue, most progress can be found in literature on dark patterns, which began just over a decade ago. In 2010, Brignull introduced the term of dark patterns as \textit{``tricks used in websites and apps that make you do things that you didn't mean to''}~\cite{brignull2015dark}. 
Throughout the past decade, many examples of dark patterns have been identified, generating a rich taxonomy of specific dark pattern types and dark pattern strategies~\cite{brignull2015dark, bosch2016, greenberg_dark_nodate, conti_malicious_2010, gray2018, Gray2020a, mathur2019, zagal_dark_2013}.
While this is a positive move in terms of better understanding unethical interface design practises, the body of work suffers somewhat from a lack of consensus, continuity, and clear definition about what exactly is being examined across various different studies. This creates a significant problem when trying to devise ways to protect users against their potentially harmful effects, including efforts to help users identify unethical design practises more easily, and attempts to regulate them.

This lack of consensus and continuity in how concepts are defined also echoes a general problem in human-machine dialogue (HMD) research, which has been highlighted in a number of recent reviews of research in the field~\cite{clark2019state, seaborn-urakami_2021, seaborn2021measuring, bruggemeier2020user}. The aim of this provocation paper is to draw attention to the potential for learning from the problems within these bodies of work and to further encourage efforts to bring clarity to how we define and measure examples of unethical design in CUIs. As a research community, we might then be able to establish more effective ways of identifying and measuring unethical design practises in the context of CUI development before they become as common and problematic as they already in graphical interface interaction.

\section{Fundamentals: Persuasion \& Manipulation}
To a certain degree persuasion is a fundamental feature of design. Ideally, the aim is to design objects that signify their potential uses and constraints so people can interact with them as intuitively as possible. That is, we look to create objects that gently nudge people toward using them in a certain way. This is a common general understanding of design in HCI research, and echoes widely known basic principles for ensuring alignment between a user’s mental model and a system forwarded by Norman in '\textit{The Design of Everyday Things}'~\cite{norman2002design}. Yet, in recent years the notion of persuasive design has begun to acquire negative connotations, largely due to a rise in design techniques aimed at governing and exploiting peoples' decision-making. Thaler and Sunstein provide an example of this type of design technique when introducing the term of \textit{Nudges}~\cite{leonard_richard_2008} to describe interventions that alternate peoples' decision making process in a predictable way, allowing design to be used to navigate a users' focus into a predefined direction or goal. However, approaches of this nature, that take advantage of our cognitive biases, can be used equally efficiently for benevolent or malevolent ends; as is evidenced by the growth in research on dark patterns
across numerous domains.

This raises questions about how we define persuasive design, and indeed, how we draw distinctions between designs regarded as persuasive versus manipulative. Nudges have certainly been used in beneficial ways, including improving peoples' eating~\cite{sarcona2017differences}, healthcare~\cite{zhao2016can}, fitness~\cite{medica2018fitness}, sleep~\cite{hosszu2019sleep} and relaxation practices~\cite{weekly2018review}. It would also be inaccurate to assume people are naive to how these systems work, with the implicit nature of persuasive cognitive approaches being both part of the appeal and part of the reason they are effective~\cite{madhusanka2021implicit}. However, they also share fundamental similarities with designs used to encourage users to make potentially harmful decisions; so-called dark patterns. They disarm a person of their autonomy, albeit temporarily, by encouraging them to make choices they might not have made otherwise. Therein lies our ethical obligation and challenges as HCI researchers: to ensure we combat unethical acts of manipulative interface design, whilst ensuring we continue to deliver persuasive interface designs that improve the lives of people who use them. We argue that establishing clear conceptual definitions is the first step toward consistent identification and measurement of unethical interface design in CUIs.

Based on established psychological definitions, we suggest drawing a distinction between persuasive design and manipulative design. Here, persuasion is defined as, \emph{“an active attempt by one person to change another person’s attitudes, beliefs, or emotions associated with some issue, person, concept, or object”}~\cite{persuasion_apa}. To bring this into context with HCI research, we suggest a slight adaptation, defining persuasive design as, \emph{‘an active attempt to influence a person’s behaviours, attitudes, beliefs, or emotions associated through interface design’}. On the other hand, manipulation is described in psychology as, \emph{‘behaviour designed to exploit, control or otherwise influence others to one’s own advantage’}~\cite{manipulation_apa}; or in the context of HCI, \emph{‘designs aimed at exploiting, controlling or otherwise influencing users to one’s own advantage’}. 
Our challenge in bringing clarity to this distinction lies in developing ways to identify and measure occurrences of unethical manipulative design, which will aid us in defending the benefits of using psychological knowledge to improve people's lives through technology also.

\section{Understanding and Measuring Unethical Design: Lessons from Dark Patterns}

\begin{table*}[t!]
\resizebox{0.95\textwidth}{!}{%
\begin{tabular}{p{0.125\linewidth}p{0.125\linewidth}p{0.125\linewidth}p{0.125\linewidth}p{0.125\linewidth}p{0.125\linewidth}p{0.125\linewidth}p{0.125\linewidth}}
\toprule
\multicolumn{1}{c}{\begin{tabular}[c]{@{}c@{}}\LARGE\textbf{Brignull}\\ \small{2010 ~\cite{brignull2015dark}}\end{tabular}} & 
\multicolumn{1}{c}{\begin{tabular}[c]{@{}c@{}}\LARGE\textbf{Conti \& Sobiesk}\\ \small{2010~\cite{conti_malicious_2010}}\end{tabular}} &
\multicolumn{1}{c}{\begin{tabular}[c]{@{}c@{}}\LARGE\textbf{Zagal et al.}\\ \small{2013~\cite{zagal_dark_2013}}\end{tabular}} & 
\multicolumn{1}{c}{\begin{tabular}[c]{@{}c@{}}\LARGE\textbf{Greenberg et al.}\\ \small{2014~\cite{greenberg_dark_nodate}}\end{tabular}} & 
\multicolumn{1}{c}{\begin{tabular}[c]{@{}c@{}}\LARGE\textbf{Bösch et al.}\\ \small{2016~\cite{bosch2016}}\end{tabular}} & 
\multicolumn{1}{c}{\begin{tabular}[c]{@{}c@{}}\LARGE\textbf{Gray et al.}\\ \small{2018~\cite{gray2018}}\end{tabular}} & 
\multicolumn{1}{c}{\begin{tabular}[c]{@{}c@{}}\LARGE\textbf{Gray et al.}\\ \small{2020~\cite{Gray2020a}}\end{tabular}} & 
\multicolumn{1}{c}{\begin{tabular}[c]{@{}c@{}}\LARGE\textbf{Mathur et al.}\\ \small{2019~\cite{mathur2019}}\end{tabular}} 
\\ \midrule

\multicolumn{1}{c|}{\begin{tabular}[t]{l}   \medskip\large\textit{· Trick Questions} \\
                                            \medskip\large\textit{· Sneak Into Basket} \\
                                            \medskip\large\textit{· Roach Motel} \\
                                            \medskip\large\textit{· Privacy Zuckering} \\
                                            \medskip\large\textit{· Confirmshaming} \\
                                            \medskip\large\textit{· Disguised Ads} \\
                                            \medskip\large\textit{\begin{tabular}[l]{@{}l@{}}· Price Comparison\\\hspace{5pt}Prevention\end{tabular}} \\
                                            \medskip\large\textit{· Misdirection} \\
                                            \medskip\large\textit{· Hidden Costs} \\
                                            \medskip\large\textit{· Bait and Switch} \\
                                            \medskip\large\textit{· Forced Continuity} \\
                                            \medskip\large\textit{· Friend Spam} \\
\end{tabular}} &                                       
\multicolumn{1}{c|}{\begin{tabular}[t]{l}   \medskip\large\textit{· Coercion} \\ 
                                            \medskip\large\textit{· Distraction} \\
                                            \medskip\large\textit{· Forced Work} \\
                                            \medskip\large\textit{\begin{tabular}[l]{@{}l@{}}· Manipulating\\\hspace{5pt}Navigation\end{tabular}} \\
                                            \medskip\large\textit{\begin{tabular}[l]{@{}l@{}}· Restricting\\\hspace{5pt}Functionality\end{tabular}} \\
                                            \medskip\large\textit{· Trick} \\
                                            \medskip\large\textit{· Confusion} \\
                                            \medskip\large\textit{· Exploiting Errors} \\
                                            \medskip\large\textit{· Interruption} \\
                                            \medskip\large\textit{· Obfuscation} \\
                                            \medskip\large\textit{· Shock} \\
\end{tabular}} &                                                           
\multicolumn{1}{c|}{\begin{tabular}[t]{l}   \medskip\large\textit{· Grinding} \\ 
                                            \medskip\large\textit{· Impersonation} \\
                                            \medskip\large\textit{· Monetized Rivalries} \\
                                            \medskip\large\textit{· Pay to Skip} \\
                                            \medskip\large\textit{\begin{tabular}[l]{@{}l@{}}· Playing by\\\hspace{5pt}Appointment\end{tabular}}\\
                                            \medskip\large\textit{\begin{tabular}[l]{@{}l@{}}· Pre-Delivered \\\hspace{5pt}Content\end{tabular}}\\
                                            \medskip\large\textit{\begin{tabular}[l]{@{}l@{}}· Social Pyramid\\\hspace{5pt}Schemes\end{tabular}}\\
\end{tabular}} &                                                                
\multicolumn{1}{c|}{\begin{tabular}[t]{l}   \medskip\large\textit{· Attention Grabber} \\ 
                                            \medskip\large\textit{· Bait and Switch} \\ 
                                            \medskip\large\textit{\begin{tabular}[l]{@{}l@{}}· The Social Network\\\hspace{5pt}Of Proxemic Contracts\\\hspace{5pt}Or Unintended\\\hspace{5pt}Relationships\end{tabular}}\\ 
                                            \medskip\large\textit{· Captive Audience} \\ 
                                            \medskip\large\textit{· We Never Forget} \\ 
                                            \medskip\large\textit{\begin{tabular}[l]{@{}l@{}}· Disguised Data\\\hspace{5pt}Collection\end{tabular}}\\ 
                                            \medskip\large\textit{\begin{tabular}[l]{@{}l@{}}· Making Personal\\\hspace{5pt}Information Public\end{tabular}}\\ 
                                            \medskip\large\textit{· The Milk Factor} \\ 

\end{tabular}} &                                                              
\multicolumn{1}{c|}{\begin{tabular}[t]{l}   \medskip\large\textit{· Privacy Zuckering} \\ 
                                            \medskip\large\textit{\begin{tabular}[l]{@{}l@{}}· Hidden Legalese\\\hspace{5pt}Stipulations\end{tabular}} \\
                                            \medskip\large\textit{· Shadow User Profiles} \\
                                            \medskip\large\textit{· Bad Defaults} \\
                                            \medskip\large\textit{· Immortal Accounts} \\
                                            \medskip\large\textit{· Information Milking} \\
                                            \medskip\large\textit{· Forced Registration} \\
                                            \medskip\large\textit{\begin{tabular}[l]{@{}l@{}}· Address Book\\\hspace{5pt}Leeching\end{tabular}} \\

\end{tabular}} &                                                           
\multicolumn{1}{c|}{\begin{tabular}[t]{l}   \medskip\large\textit{· Nagging} \\ 
                                            \medskip\large\textit{· Obstruction} \\
                                            \medskip\large\textit{· Sneaking} \\
                                            \medskip\large\textit{· Interface Interference} \\
                                            \medskip\large\textit{· Forced Action} \\

\end{tabular}} &                                                        
\multicolumn{1}{c|}{\begin{tabular}[t]{l}   \medskip\large\textit{· Automating the User} \\ 
                                            \medskip\large\textit{· Two-Faced} \\
                                            \medskip\large\textit{· Controlling} \\
                                            \medskip\large\textit{· Entrapping} \\
                                            \medskip\large\textit{· Nickling-And-Diming} \\
                                            \medskip\large\textit{· Misrepresenting} \\

\end{tabular}} &  
\multicolumn{1}{l}{\begin{tabular}[t]{l}    \medskip\large\textit{· Countdown Timers} \\ 
                                            \medskip\large\textit{\begin{tabular}[l]{@{}c@{}}· Limited-time\\\hspace{5pt}Messages\end{tabular}} \\
                                            \medskip\large\textit{\begin{tabular}[l]{@{}c@{}}· High-demand\\\hspace{5pt}Messages\end{tabular}} \\
                                            \medskip\large\textit{· Activity Notifications} \\
                                            \medskip\large\textit{· Confirmshaming} \\
                                             \medskip\large\textit{\begin{tabular}[l]{@{}c@{}}· Testimonials\\\hspace{5pt}of Uncertain\\\hspace{-18pt}Origins\end{tabular}} \\
                                            \medskip\large\textit{· Hard to Cancel} \\
                                            \medskip\large\textit{· Visual Interference} \\
                                            \medskip\large\textit{· Low-stock Messages} \\
                                            \medskip\large\textit{· Hidden Subscriptions} \\
                                            \medskip\large\textit{· Pressured Selling} \\
                                            \medskip\large\textit{· Forced Enrollment} \\
\end{tabular}} 

\\ \bottomrule
\end{tabular}
}
\caption{This table lists 62 types of dark pattern described in prior research.}
\Description[This table shows types of dark patterns described in prior research.]{This table presents a total of 62 types of dark patterns divided by their original publication over 8 columns. These columns are as follows: 
1. Brignull et al. 
· Trick Questions
· Sneak Into Basket
· Roach Motel
· Privacy Zuckering
· Confirmshaming
· Disguised Ads
· Price Comparison
Prevention
· Misdirection
· Hidden Costs
· Bait and Switch
· Forced Continuity
· Friend Spam 

2. Conti & Sobiesk
· Coercion
· Distraction
· Forced Work
· Manipulating
Navigation
· Restricting
Functionality
· Trick
· Confusion
· Exploiting Errors
· Interruption
· Obfuscation
· Shock

3. Zagal et al.
· Grinding
· Impersonation
· Monetized Rivalries
· Pay to Skip
· Playing by
Appointment
· Pre-Delivered
Content
· Social Pyramid
Schemes

4. Greenberg et al.
· Attention Grabber
· Bait and Switch
· The Social Network
Of Proxemic Contracts
Or Unintended
Relationships
· Captive Audience
· We Never Forget
· Disguised Data
Collection
· Making Personal
Information Public
· The Milk Factor

5. Bösch et al.
· Privacy Zuckering
· Hidden Legalese
Stipulations
· Shadow User Profiles
· Bad Defaults
· Immortal Accounts
· Information Milking
· Forced Registration
· Address Book
Leeching

6. Gray et al. 
· Nagging
· Obstruction
· Sneaking
· Interface Interference
· Forced Action

7. Gray et al.
· Automating the User
· Two-Faced
· Controlling
· Entrapping
· Nickling-And-Diming
· Misrepresenting

8. Mathur et al.
· Countdown Timers
· Limited-time
Messages
· High-demand
Messages
· Activity Notifications
· Confirmshaming
· Testimonials
of Uncertain
Origins
· Hard to Cancel
· Visual Interference
· Low-stock Messages
· Hidden Subscriptions
· Pressured Selling
· Forced Enrollment
}
\label{tab:dp_taxonomy}
\end{table*}

Research into unethical interface design practises began with examinations of e-commerce websites by Brignull~\cite{brignull2015dark}. The work identified twelve specific examples of unethical design aimed at inhibiting people’s ability to make informed choices. These include, \textit{Sneak into Basket}, \textit{Hidden Costs}, and \textit{Price Comparison Prevention}, which all operate on the premise of obscuring information and potentially misleading users into buying unwanted or unnecessarily expensive products. Other dark patterns defined by the author include, \textit{Forced Continuity}, \textit{Privacy Zuckering}, and \textit{Roach Motel}, which are all tactics aimed at forcing people to sign-up, or stay signed up for accounts and services they might not require anymore. These examples are also used by service providers to gain access to private data without fully informing users why it is being collected, who has access to it, or even what it might be used for. Since then, researchers have described and defined a plethora of other examples, with a recent review of the literature from Mathur et al.~\cite{Mathur2021} identifying 62 specific types of dark patterns. 
For readability, Table~\ref{tab:dp_taxonomy} offers a complete overview.
Understanding how these existing dark patterns were established and how they are used to facilitate unethical design in graphical user interfaces (GUIs) could allow us to prevent similar developments in CUIs.


While it may seem like unethical GUI design has been a perennial problem for HCI researchers, literature on dark patterns shows that these tactics and the form they take develop and change over time through conscious efforts made by interface design practitioners. Grey et al.~\cite{gray2018, gray_ethical_2019} identified multiple dark patterns as well as design constraints of practitioners which lead to their creation. In a first study, the authors analysed an image-based corpus to define five types of dark patterns that practitioners engage in when developing manipulative designs~\cite{gray2018}. Most of Gray at al.'s dark patterns include descriptions from prior works. For example, the \textit{Obstruction} dark pattern, used to make processes unnecessarily difficult, incorporates Brignull's \textit{Roach Motel}, \textit{Price Comparison Prevention}, and \textit{Intermediate Currency}~\cite{brignull2015dark}. In a follow up to this work, Gray et al. also analysed 4775 user-generated posts of the Reddit sub-forum \textit{r/assholedesign}~\cite{gray_ethical_2019}. Analysis resulted in identifying six properties of ``asshole designers''. The work is particularly useful for understanding the origins of dark patterns and how they emerge from constraints under which practitioners work.

While many of the previously described dark patterns are applicable to different domains, research has also found domain specific examples. For example, Zagal et al. identified seven dark patterns that related to video game mechanics~\cite{zagal_dark_2013}. While certain patterns exploit a game's ecosystem of connected users, such as \textit{Social Pyramid Schemes} and \textit{Impersonation}, others impact game-play experience like \textit{Grinding} and \textit{Playing by Appointment}.
Elsewhere, Greenberg et al.~\cite{greenberg_dark_nodate} consider dark patterns in conjunction with proxemics theory~\cite{hall_hidden_1966}. Identifying nine types of dark pattern in total, the authors discuss interactions with manipulative design in spatial environments. The \textit{Attention Grabber} and \textit{Disguised Data Collection} dark patterns, for instance, could be used in the design of digital billboards that exploit people's proximity and personal data to deliver personalised advertising.

Understanding the creation of dark patterns and analysing their occurrences helps to close the ``cultural lag''~\cite{ogburn_social_1922} where the creation of ethical guidelines inevitably lags behind the release of novel dark patterns and even novel technologies. However, recent attempts by the authors to apply the aforementioned corpus of dark patterns in the domain of social media highlight a central problem in this body of work: it is hallmarked by a high degree of overlap between definitions, inconsistent terminology, and descriptions that operate on different levels often without explicit acknowledgement. That is, the difference between specific dark pattern designs and broader dark pattern strategies is not always recognised. 
Attempts to apply taxonomies of dark patterns also shows that while some established dark patterns types are applicable to other domains, it is also likely that novel CUI specific dark patterns types and strategies will need to be described and defined. Further, while this is particularly problematic from a research perspective, we see an even more urgent need to begin this work so we might also better protect users. 


\section{Ways to Combat Unethical Interface Design}
With regards to technology, particularly online interfaces, studies have shown that the burden to counteract dark patterns often falls on users~\cite{digeronimo2020, BongardBlanchy2021}. Although regulations, such as the GDPR~\cite{eu_gdpr_2016} or the CCPA~\cite{ccpa_2018}, aim to protect users in online environments, the previously mentioned cultural lag~\cite{ogburn_social_1922} means these efforts are struggling to counter all problematic designs described under the umbrella of manipulative design. Indeed, one could argue they have led to the creation of new dark patterns. The design of cookie consent banners that favour 'accept all' options over offering users greater control, shows how design can be used to easily negate efforts to combat dark patterns, and how current regulatory efforts fail to address the fundamental nature of manipulative design. This has lead to designs exploiting cognitive biases not covered by the GDPR, such as anchoring effects, to steer users into giving consent that they might not have given were they provided with a neutral choice~\cite{Gray2021, mathur2019}. 

In January 2022, the European Union proposed a new article 13a, as part of the Digital Service Act~\cite{eu_dsa_2022}, to address previous concerns. Offering a generalised definition of problematic design that is closely worded to Mathur et al's.~\cite{Mathur2021} definition of dark patterns, article 13a contains an extendable list of specific interface designs to be regulated. However, we argue that this approach may lead to a similar problem seen in attempts to apply dark pattern taxonomies across different interfaces and domains. That is, the taxonomies quickly become very large, difficult to maintain, and not always appropriate across interfaces types and use cases.
Further, creating an ever growing list of examples may also deepen this problem over-complicating crucial efforts to combat unethical design.




\begin{table}[ht]
\renewcommand{\arraystretch}{1.4}
\begin{tabular}{p{0.22\linewidth}p{0.68\linewidth}}
\toprule
\multicolumn{2}{c}{{\begin{tabular}[c]{@{}c@{}}\textbf{Mathur et al. 2019 \cite{mathur2019}}\\ Dark Pattern Characteristics\end{tabular}}}                                \\ \midrule
Characteristic  & Question                                                                                                                          \\ \midrule
\cellcolor[gray]{0.95}Asymmetric          & \cellcolor[gray]{0.95}Does the user interface design impose unequal weights or burdens on the available choices presented to the user in the interface? \\
Covert              & Is the effect of the user interface design choice hidden from the user? \\
\cellcolor[gray]{0.95}Deceptive           & \cellcolor[gray]{0.95}Does the user interface design induce false beliefs either through affirmative misstatements, misleading statements, or omissions?\\
Hides Information  & Does the user interface obscure or delay the presentation of necessary information to the user? \\
\cellcolor[gray]{0.95}Restrictive         & \cellcolor[gray]{0.95}Does the user interface restrict the set of choices available to users?  \\ 
\bottomrule
\end{tabular}
\caption{This table lists the introductory questions Mathur et al. (2019)~\cite{mathur2019} gave for each dark pattern characteristic.}
\Description[Dark Pattern Characteristics]{This table lists the introductory questions Mathur et al. (2019) gave for each dark pattern attribute. The attributes and their introductory questions are as follows: 
1. Asymmetric: Does the user interface design impose unequal weights or burdens on the available choices presented to the user in the interface?

2. Covert: Is the effect of the user interface design choice hidden from the user?

3. Deceptive: Does the user interface restrict the set of choices available to users?

4. Hides Information: Does the user interface design induce false beliefs either through affirmative misstatements,
misleading statements, or omissions?

5. Restrictive: Does the user interface obscure or delay the presentation of necessary information
to the user?}
\Description[Dark Pattern Characteristics]{This table lists the introductory questions Mathur et al. (2019) gave for each dark pattern attribute. The attributes and their introductory questions are as follows: 
1. Asymmetric: Does the user interface design impose unequal weights or burdens on the available choices presented to the user in the interface?

2. Covert: Is the effect of the user interface design choice hidden from the user?

3. Deceptive: Does the user interface restrict the set of choices available to users?

4. Hides Information: Does the user interface design induce false beliefs either through affirmative misstatements,
misleading statements, or omissions?

5. Restrictive: Does the user interface obscure or delay the presentation of necessary information
to the user?}
\label{tab:mathur2019}
\end{table}

\section{Levelling up to Unethical Interface Design}
By categorising specific dark pattern types into five broad characteristics, Mathur et al.~\cite{mathur2019} offer an alternative and potential useful avenue for combating manipulative designs. This higher level categorisation is based on the cognitive biases specific designs are developed to exploit. By relying on more fundamental concepts, which are much less interchangeable than specific interface layouts, the approach offers a way to understand manipulative design that focuses on their impact on users, in a broadly applicable fashion, whilst remaining agnostic to the interface or domain in question. By being based on cognitive biases, instead of mere dark pattern definitions that stem from GUI domains, this model allows enhanced evaluation outside its original scope, such as CUIs.

\begin{figure*}[ht!]
    \centering
    \includegraphics[width=0.9\textwidth]{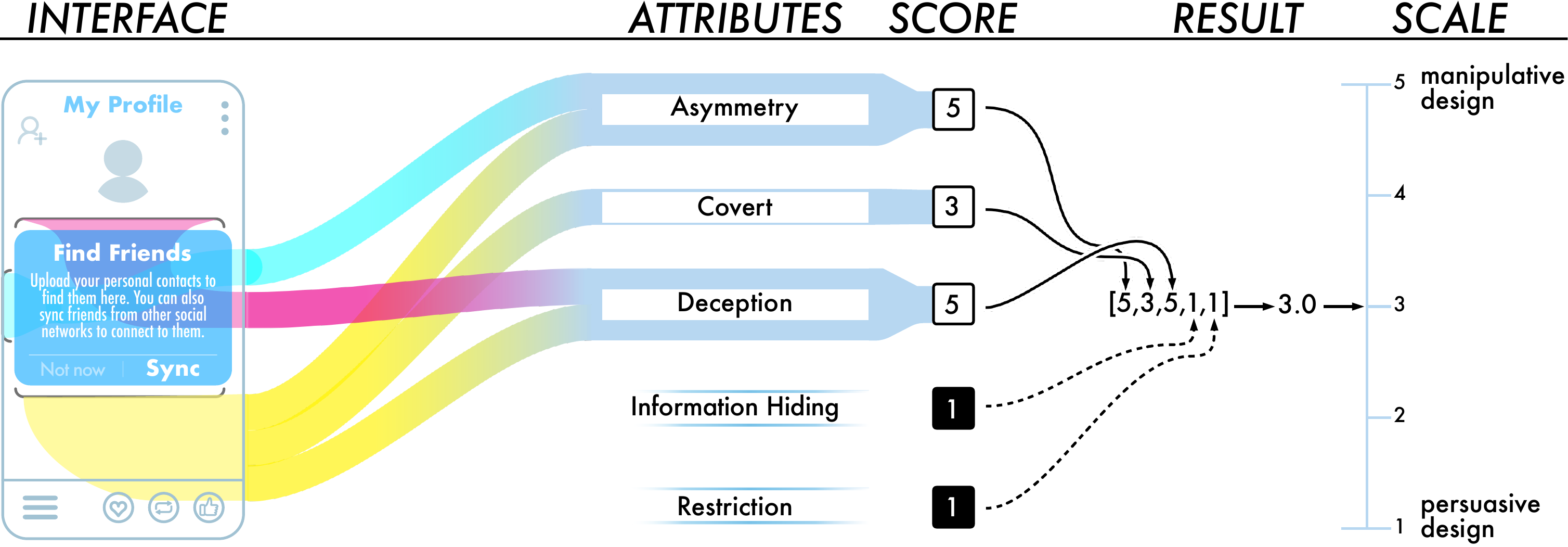}
    \caption{This figure demonstrates the derivation of a five-dimensional dark pattern vector, which can further be reduced into a single digit value, inspired by Mathur et al.'s five dark pattern characteristics and attributes~\cite{mathur2019}. Although the example is based on a screenshot of a mobile interface, it could be easily adapted for any conversational user interface.}
    \label{fig:dp_vectorisation}
    \Description[This figure demonstrates the derivation of a five-dimensional dark pattern vector, which can further be reduced into a single digit value, inspired by Mathur et al.'s five dark pattern characteristics and attributes~\cite{mathur2019}. Although the example is based on a screenshot of a mobile interface, it could be easily adapted for any conversational user interface.]{The figure shows a five step procedure to get a dark pattern score based on Mathur et al.'s five dark pattern characteristics. The first step shows an illustration of a screenshot containing multiple dark patterns. In the second step, these dark patterns are linked to each applying attribute respectively. In this exemplar, the attributes asymmetry, covert, and deception are recognised. The attributes information hiding and restriction are not. Steps four and five are then deriving a score for }
\end{figure*}

In an attempt to advance their model and differentiate between manipulative and persuasive designs (i.e, between dark patterns and bright patterns), we developed a technique that allows us to evaluate individual examples based on characteristics that stem from the cognitive biases they target. We therefore consider each of the five characteristic by asking a specific question to assess impact on each of these dimensions, as seen in Table~\ref{tab:mathur2019}. By assigning each a value from 1 (not at all) to 5 (extremely), we are able to evaluate persuasive designs across the five dimensions, indicating the degree to which they might be regarded as persuasive or manipulative. The benefit of this approach is that we gain an overall score that can be used to determine the degree to which a specific design is either persuasive or manipulative, whilst identifying which dimensions of persuasion a manipulative design targets. Depending on the context and situation in which a design is evaluated, the score determining the degree of persuasive versus manipulative design can be adapted by alternating the threshold set to identify what is acceptable design depending on the score. Figure~\ref{fig:dp_vectorisation} visualises the steps of this process. By providing this clearly defined and measurable conceptualisation of persuasive and manipulative design, this model yields a certain duality. On one hand, we might help regulators combat unethical interface design practices in a consistent and broadly applicable fashion, whilst protecting the benefits of persuasion. On the other, this approach could allow practitioners to evaluate their own designs through user studies.


\section{Current Caveats To Keep In Mind}
This provocation paper addresses unethical design in CUIs but discusses the topic with the means of manipulation and exploitation of cognitive biases. We understand a distinction between ethical design and measurement of unethical practices. Yet, we argue that by learning which unethical practices are at play, we are able to compare and understand practitioners' strategies better while promoting more conscious handling of design techniques that, in the wrong context, may exploit cognitive biases harming the user. The currently available amount of tools to assess the good or bad in design is limited while a growing demand to evaluate interfaces ethically can be seen across disciplines. By utilising knowledge about cognitive biases, and exploitation thereof, we are able to understand malicious interface strategies better and can further classify them to share new knowledge between research communities. As a basis for this research, cognitive biases describe basic behaviour traits, certain strategies and heuristics under which decisions are formed, shared among all humans. Building on this existing knowledge, we aim to establish a robust measurement that also allows for comparison of interfaces. 

We acknowledge the early stage of this endeavour and are aware of current limitations. Arguably, a numeric and finite score as a determiner for how ethical an interface is may be appealing to different cohorts, whether in the context of regulation or user interface design. As all ordinal scales, however, the proposed approach can only represent a limited abstraction of manipulative dimensions in an interface. Moreover, the questionnaire has not yet been verified and thus it is uncertain how effective it will prove to be in action. In this early stage, we rely on the five characteristics proposed by Mathur et al.~\cite{mathur2019}. By rooting their characteristics in cognitive bias research, they gain the advantage of being similarly effective across domains. Still, it is questionable whether these exact five dimensions are as effective in the context of CUIs when compared to their origin in GUIs. This could easily be addressed in future research by investigating the differences and unanimity between cognitive biases exploited throughout different kinds of interface modalities. Further studies could then look at the variety of described cognitive biases to identify exploitation across domains to offer alternative sets of questions and target each modality precisely. For example, where colourisation of certain buttons may promote some choices over others in GUI contexts, in voice based CUIs colour of voice, emotional speech, and pitch could be misused for similar deceptions.

\section{Conclusion}
In this provocation piece, we argue that current attempts to account for manipulative interface design are curtailed by a lack of clarity and continuity around how concepts are defined; hampering efforts to measure and combat their use. We also argue that there are valuable lessons to learn from previous work on GUI dark patterns that might help us head off these problems in the realm of CUIs. Indeed, the approach we suggest may prove useful across multiple domains and interface contexts, and stands to benefit researchers, regulators and potentially users also. Further, by aligning the conceptualisation with how manipulative designs function, rather than how they manifest, we hope to make it much more difficult for designers to circumnavigate. Manipulative design that targets our cognitive biases represents a real and pertinent danger to people, and difficulties faced, even by knowledgeable users and regulators, highlight an urgent need to develop quantifiable approaches that can be easily understood by a range of stakeholders involved in the fight against unethical interface design practices.


\begin{acks}
The research of this project was partially supported by the Klaus Tschira Stiftung gGmbH. 
\end{acks}
\bibliographystyle{ACM-Reference-Format}
\bibliography{references.bib}


\end{document}